\documentclass[lettersize,journal]{IEEEtran}
\usepackage{xcolor}
\usepackage{amsmath,amsfonts}
\usepackage{algorithmic}
\usepackage{algorithm}
\usepackage{array}
\usepackage[caption=false,font=normalsize,labelfont=sf,textfont=sf]{subfig}
\usepackage{textcomp}
\usepackage{stfloats}
\usepackage{url}
\usepackage{verbatim}
\usepackage{graphicx}
\usepackage{cite}
\begin{document}

\title{On-chip arbitrary dispersion engineering with deep photonic networks}

\author{Kazim Gorgulu$^{1, 2}$, Aycan Deniz Vit$^{1}$, Ali Najjar Amiri$^{3}$, \\ Emir Salih Magden$^{1,4*}$%
\\
$^{1}$\IEEEmembership{Department of Electrical \& Electronics Engineering, Koç University, 34450, Istanbul, Turkey}\\
$^{2}$\IEEEmembership{Sicoya GmbH, 12489 Berlin, Germany}\\
$^{3}$\IEEEmembership{Department of Electrical and Computer Engineering, Northwestern University, Evanston, IL 60208, USA}
$^{4}$\IEEEmembership{KUIS-AI Center, Koc University, 34450, Istanbul, Turkey}\\%
\thanks{*Corresponding author: esmagden@ku.edu.tr}}



\maketitle

\begin{abstract}
We demonstrate the design and optimization of on-chip arbitrary dispersion profiles using deep photonic networks constructed from custom-designed Mach-Zehnder interferometers. These photonic networks employ optimizable waveguide tapers, enabling precise engineering of wavelength-dependent dispersion profiles. We experimentally demonstrate a proof-of-concept two-port photonic network that exhibits a highly nonintuitive triangular dispersion profile over the wavelength range of 1.54 µm to 1.58 µm while simultaneously achieving a flat-band transmission with an insertion loss of less than 0.7 dB. We also illustrate the potential of multi-port photonic networks to enhance design freedom for more complex and customizable dispersion profiles, enabling new possibilities for on-chip dispersion engineering.
\end{abstract}


\section{Introduction}

Efficient dispersion management is crucial for applications such as high-speed optical communication and optical signal processing \cite{chenchip, dolgaleva2011integrated}. Previously, on-chip dispersion control has been explored using micro-ring resonators \cite{madsen1999integrated}, photonic crystals \cite{vlasov2005active}, arrayed waveguide gratings \cite{wang2020integrated}, and Bragg gratings \cite{sahin2017large}. Most of these efforts have targeted achieving a constant dispersion within a given band for compensation of undesired dispersion in communication systems. However, precise dispersion control is also essential for applications such as pulse shaping and quantum information processing \cite{weiner2011ultrafast, mower2013high}, where devices with arbitrarily defined dispersion profiles enabled by a large number of optimizable design parameters are critical.

Early pulse shaping relied on free-space diffraction gratings and spatial light modulators \cite{weiner2011ultrafast}, while on-chip implementations used cascaded tunable ring resonators for improved pulse profiles \cite{madsen1998optical}. In parallel, Notaros et al. demonstrated the critical role of precise dispersion control in quantum information processing with a tunable ring resonator array offering 30 degrees of freedom, enabling photonic quantum data locking \cite{notaros2017programmable}.
Similar approaches using arrayed waveguide gratings and multi-stage Mach-Zehnder interferometers (MZIs) enable on-chip dynamic dispersion engineering \cite{tahvili2013inp, moreira2016programmable}. However, additional stages increase complexity, and achieving both custom dispersion and low-loss, flat-band transmission remains challenging, with reported losses up to 20 dB.

In recent years, MZI-based photonic networks have gained significant attention for their low-loss performance and design flexibility, enabling applications in photonic signal processing and the realization of arbitrary optical responses \cite{bogaerts2020programmable, ashtiani2022chip, najjar2024deep}. Leveraging these capabilities, in this work, we present the design of on-chip deep photonic networks incorporating custom MZIs to engineer dispersion profiles with arbitrary specifications. In these networks, each MZI implements custom phase delays through optimizable waveguide tapers, allowing controllable dispersion profiles across a wide spectrum of interest. To design these networks, we develop a custom simulation and optimization framework that simultaneously engineers both dispersion and transmission profiles. Using this framework, we create and experimentally demonstrate a proof-of-concept photonic network on a standard 220-nm silicon-on-insulator platform, featuring a highly nonintuitive triangular-shaped dispersion profile over the wavelength range of 1.54 µm to 1.58 µm, and achieving flat-band transmission with less than 0.7 dB insertion loss. Furthermore, we demonstrate more complex capabilities of multi-port photonic networks for realizing highly customized dispersion profiles.

\section{Results and Discussion}

We illustrate the general structure and working principle of our networks in Fig 1. Fig. 1a shows a two-input, two-output integrated photonic network consisting of several cascaded MZI units. Each MZI performs a 2×2 linear transformation, denoted by \( T_i\). Fig. 1b illustrates a schematic view of one of these MZIs, composed of two directional couplers and two pairs of custom tapers. For our demonstrations, we use identical directional couplers throughout the network, each designed as a 50/50 splitter at 1550 nm. In contrast, each custom taper is designed with a set of optimizable widths, as shown in Fig. 1c. Unlike propagation in a straight waveguide with a fixed spectral dispersion profile, these custom tapers enable tailored wavelength-dependent dispersion profiles due to their varying width along the propagation path. To illustrate this, Fig. 1d presents the group delay spectrum of a 10 µm-long custom silicon taper, with widths ranging from 400 nm to 500 nm. For comparison, the group delays of three different 10 µm-long straight waveguides with uniform widths of 445 nm, 450 nm, and 455 nm are also plotted. The slope of the custom taper’s group delay clearly differs from that of the straight waveguides, demonstrating that the custom taper achieves a unique delay spectrum that cannot be replicated by straight waveguides. In a typical photonic network, many such custom tapers are used, and the dependence of spectral dispersion on waveguide geometry ultimately enables the realization of custom, application-specific phase and dispersion profiles.

\begin{figure}[t]
\centering
\includegraphics[width=\linewidth]{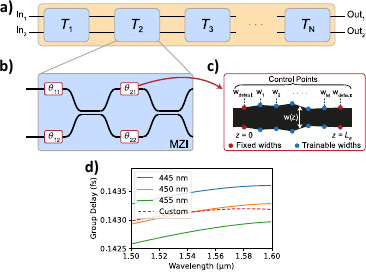}
\caption{a) Photonic network architecture with cascaded Mach-Zehnder interferometers (MZIs). b) Single MZI with two directional couplers and two pairs of custom tapers for specific spectral phase profiles. c) Custom waveguide with trainable width parameters. d) Group delay spectrum of a 10 µm-long custom silicon taper (widths 400–500 nm) compared to three straight silicon waveguides (widths 445 nm, 450 nm, 455 nm).}
\label{fig:fig1}
\end{figure}

We model the MZI networks using a transfer matrix formulation. The transfer matrix of a directional coupler is given by

\begin{equation}
C = e^{-j\varphi(\lambda)} 
\begin{bmatrix}
t(\lambda) & -jq(\lambda) \\ 
-jq(\lambda) & t(\lambda)
\end{bmatrix}
\label{eq:refname1}
\end{equation}

\noindent where $t$ and $q$ are through-coupling and cross-coupling coefficients, respectively, and $\varphi$ is the phase accumulated in the coupler. The transfer matrix for each pair of tapers is given by

\begin{equation}
P_k =
\begin{bmatrix}
e^{-j\theta_{k1}(\lambda)} & 0 \\ 
0 & e^{-j\theta_{k2}(\lambda)}
\end{bmatrix}
\end{equation}

\noindent where \( \theta_{k1} \) and \( \theta_{k2} \) are the accumulated phases in the top and bottom arms, respectively. The accumulated phase in these waveguides is given by  
\( \theta(\lambda) = \int_{0}^{L} \frac{2\pi}{\lambda} n_{\text{eff}}(w(z),\lambda) \, dz \),  
where \( n_{\text{eff}} \) is the effective index of the waveguide as a function of wavelength and width, and \( L \) is the length of the custom waveguide. The width profile \( w(z) \) is obtained by interpolating the optimizable widths w$_1$ through w$_M$. Finally, the overall linear transfer matrix of one MZI unit is given by  
\( T = C P_2 C P_1 \), where \( C\) denotes the directional couplers and \( P_1\) and \( P_2\) denote phase delay sections.

In our designs, while all directional couplers in the MZI network are identical, each phase delay section is uniquely optimized to achieve the desired optical response. To enable fast simulation and optimization for arbitrarily defined optical functions, we developed an automatic differentiation-compatible framework leveraging existing machine learning libraries \cite{bradbury2018jax, trax}. The design and optimization flow for an example photonic network is shown in Fig. 2a. The first step is a forward propagation operation, projecting light from input to output waveguides using the transfer matrix calculations described above. For this, forward propagator uses the coupling coefficients $t$, $q$, and phase shift $\varphi$ for the directional couplers, along with the effective index $n_{eff}$ for calculating \( \theta(\lambda)\) in the custom phase delay sections. The directional coupler coefficients are interpolated from precomputed 3D finite-difference-time domain simulations, and the effective indices are extracted from finite-difference eigenmode calculations using the Silicon Photonics Toolkit package \cite{vit2023silicon}.

To enable gradient-based dispersion optimization, all parameters are modeled as twice continuously differentiable with respect to wavelength, and effective indices as differentiable with respect to width. At the network output, the complex signal is separated into phase and amplitude components to compute dispersion $D_c(\lambda)$ and transmission $T_c(\lambda)$. A loss function $L(\mathbf{W})$ is defined as the difference between the calculated and target responses, summed over the spectrum of interest, where $\mathbf{W}$ represents all trainable width parameters. A weighting factor $\alpha$ controls the relative contributions of dispersion and transmission terms. Finally, the gradient of $L(\mathbf{W})$ with respect to widths is computed through back-propagation and used to update the custom taper widths in each phase delay section. This iterative process is repeated until $L(\mathbf{W})$ is sufficiently minimized.

To prevent the optimizable widths from becoming extremely wide or narrow, or undergoing abrupt consecutive changes during optimization, we implemented two regularization schemes. The first scheme maintains the taper widths close to a user-defined reference width \( w_{\text{ref}} = 450\,\text{nm} \), ensuring single-mode operation and sufficient modal confinement by avoiding excessively wide or narrow widths. For a given custom taper, this is achieved by minimizing the error function \( R_1 = \sum_{l=1}^M (w_l - w_{\text{ref}})^2 \). The second scheme minimizes the variation between consecutive widths in a given custom taper, reducing the risk of propagation loss due to higher-order mode excitation, and is calculated as \( R_2 = \sum_{l=1}^{M-1} (w_l - w_{l+1})^2 \). The total regularization contribution for a custom taper is defined as \( R = \gamma_1 R_1 + \gamma_2 R_2 \), where \( \gamma_1 \) and \( \gamma_2 \) are hyperparameters that control the strength of each regularization term. For all custom tapers, these regularization terms are accumulated and incorporated as additional contributions to \( L(\mathbf{W}) \), creating a trade-off between design compatibility and flexibility.

\begin{figure}[t]
\centering
\includegraphics[width=\linewidth]{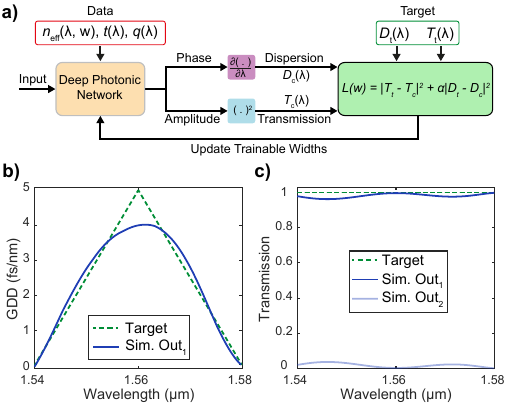}
\caption{a) Optimization cycle of photonic network. $D_c(\lambda)$ and $T_c(\lambda)$ are the calculated dispersion and transmission, respectively. $D_t(\lambda)$ and $T_t(\lambda)$ are the target dispersion and transmission, respectively. $L(\mathbf{W})$ is the loss function, and $(\alpha)$ is the weighting parameter balancing dispersion and transmission. b) Target and optimized group delay dispersion for a 2×15 network. c) Target and optimized transmittance of the same network.}
\label{fig:fig2}
\end{figure}

The proposed network architecture and optimization procedure enable the design of application-specific photonic networks with arbitrarily defined, broadband dispersion and transmission profiles. As a proof of principle, we demonstrate a photonic network that achieves an unconventional, triangular group delay dispersion profile (GDD). The target dispersion profile from 1.54 µm to 1.58 µm is shown by the dashed line in Fig. 2b. For a photonic network, the number of MZI layers is a crucial parameter, as it determines the degrees of freedom and how accurately a target dispersion profile can be realized. For this design, a 15-stage MZI network is used, with a total length of 1.08 mm. Each MZI contains four optimizable waveguide tapers, resulting in 60 tapers and 300 trainable parameters in total. A broadband input is applied at the input, and the response at the selected output (Out$_1$) is optimized for both dispersion and transmission using a multi-objective optimizer. The optimization, performed over 3000 iterations, is completed in under five minutes on a standard computer. The calculated dispersion (blue curve, Fig. 2b) closely matches the target, validating the effectiveness of the proposed approach. In addition to dispersion shaping, the simulated transmission (blue curve, Fig. 2c) closely follows the target, achieving a near-lossless response with simulated losses below 0.16 dB across the entire spectrum. The remaining power is directed to the second output port and is effectively minimized.

This photonic network was fabricated using 193 nm CMOS photolithography on 220 nm SOI wafers via imec’s multi-project-wafer foundry service. Dispersion characterization was performed using an interferometric technique based on transmission fringe observation \cite{vlasov2005active}. As shown in Fig. 3a, the device under test was placed in one arm of an integrated interferometer. Transverse electric polarized light was coupled on-chip using a tunable laser and a manual 3-paddle polarization controller, with foundry-provided edge couplers for light coupling. 2×2 on-chip power splitters/combiners were used at both the input and output of the interferometer.

\begin{figure}[t]
\centering
\includegraphics[width=\linewidth]{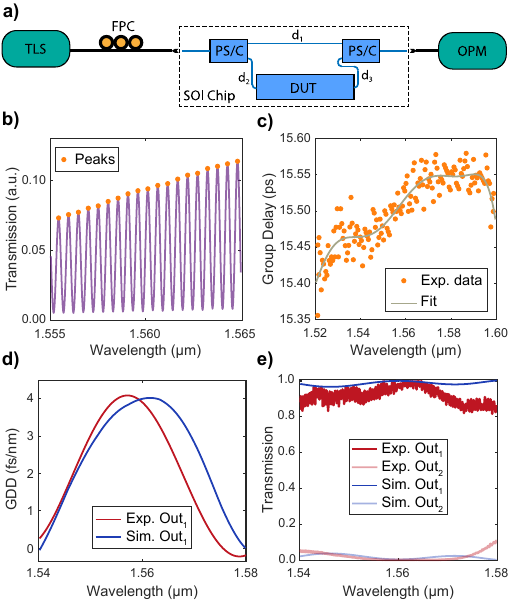}
\caption{Fig. 3. a) Illustration of the interference-based dispersion measurement setup. TLS: tunable laser source, FPC: fiber polarization controller, PS/C: power splitter/combiner, DUT: device under test, OPM: optical power meter. \(d_1\), \(d_2\), and \(d_3\) are the lengths of waveguides connecting components of the interferometer. \(d_1\)=\(d_2\)+\(d_3\) is set to eliminate any unbalanced dispersion effects through these waveguides. b) Measured interference fringes and corresponding peaks of the two-port network optimized for the triangular dispersion profile. c) Group delay extracted from interference fringes. d) Experimental dispersion derived from group-delay measurements, compared with simulated dispersion. e) Measured transmission responses of the network, compared with simulations.
}
\label{fig:fig3}
\end{figure}

Measured interference fringes and detected peaks are shown in Fig. 3b. Group delay is calculated from these peaks as \( \tau = \frac{\lambda^2}{\Delta\lambda c} \), where \( \Delta\lambda \) is the spectral period of the interference fringes, and \( c \) is the speed of light. This experimentally obtained group delay is shown in Fig. 3c, along with a polynomial fit. Dispersion is then calculated as \( \text{GDD} = \frac{d\tau}{d\lambda} \), where differentiation is performed on the polynomial-fit group delay with respect to wavelength. This dispersion result is plotted in Fig. 3d (red), in agreement with the simulated result from Fig. 2b (replotted here in blue for reference). While the experimental dispersion is slightly narrower with a blue shift, the overall profile and peak dispersion closely match the simulated data. These spectral deviations can be attributed to fabrication-induced variations in directional couplers and the taper geometries, which can shift coupling ratios, alter the intended phase response, and influence the network’s exact behavior. In deep photonic networks composed of many MZIs, these local perturbations can accumulate and influence the spectral position and shape of the dispersion profile. However, despite the presence of 15 cascaded MZI layers, the measured dispersion remains close to the designed target, indicating that the overall network behavior remains robust to the typical photolithography-induced variations in waveguide geometry.

The transmission measurements were performed using an independent replica of the same device, placed separately on the same chip. The output powers at ports Out$_1$ and Out$_2$ were measured with an optical power meter and plotted in Fig. 3e. Simulated transmission results from Fig. 2c are also replotted here for comparison. As predicted by our transfer matrix simulations, the vast majority of the output power was recorded at Out$_1$ across the measured spectrum. The measured insertion loss was less than 0.7 dB within the 1.54-1.58 µm wavelength range. The slightly higher experimental loss compared to simulations can be attributed to propagation losses due to scattering. Additionally, fabrication-induced coupling and phase deviations in the components of the MZIs can redirect power to the complementary port instead of the intended target port.

The design flexibility of our networks allows for a wide range of dispersion profiles, with more complex profiles achievable by increasing the number of layers or ports. For example, Fig. 4a shows a wider 3-port network optimized to demonstrate additional dispersion capabilities. This architecture allows the optical input to take multiple different paths depending on the wavelength, recombining at the network’s end for more flexible, independent control of spectral features. As a proof-of-concept, we optimize a step-like dispersion profile with a 3-port, 8-layer, 864-µm-long network, with results shown in Fig. 4b. The optimized MZI network achieves a broadband dispersion profile matching the step-like target across the 1.54-1.58-µm range. Using a similar 3-port topology, we also optimize for a much greater dispersion target (50 fs/nm) within a narrower 2 nm bandwidth, as shown in Fig. 4c. This 24-layer, 2.16-mm-long network achieves a near-flat dispersion profile within the bandwidth, as plotted by the blue curve. For comparison, the orange curve shows the dispersion from a 2.16-mm-long, straight, 450-nm-wide waveguide. In this case, our optimized network achieves up to 70x more dispersion within the 2 nm bandwidth than a straight waveguide of equal length. This comparison demonstrates the ability of the deep photonic network to achieve custom dispersion profiles that differ significantly from straight waveguides.

\begin{figure}[t]
\centering
\includegraphics[width=\linewidth]{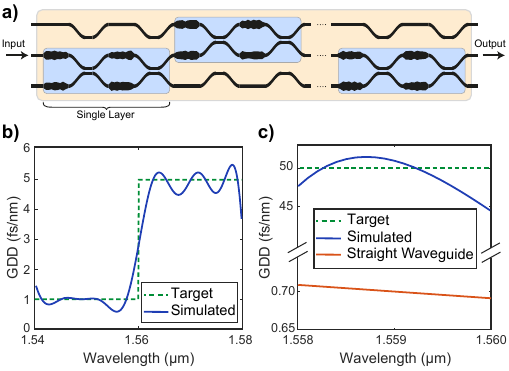}
\caption{a) Schematic of the 3-port photonic network architecture. b) Dispersion profile of a 3-port network optimized for a step-like dispersion. c) Dispersion profile of a 3-port network optimized for high dispersion, with a straight waveguide of the same length shown for comparison.}
\label{fig:fig4}
\end{figure}

The deep photonic networks presented in this work enable highly customizable dispersion profiles, going beyond what is typically achievable with ring resonators, photonic crystals, and arrayed waveguide gratings \cite{notaros2017programmable, tahvili2013inp, moreira2016programmable}. Moreover, our device exhibits significantly lower loss than these approaches, which often report insertion losses of several dB or more; in contrast, our measured loss is only 0.7 dB. Even though the maximum dispersion strength demonstrated experimentally in our current designs is lower than in some alternatives, this limitation is not fundamental, as illustrated by the simulation results in Fig. 4. Additionally, our approach can benefit from incorporating variable taper lengths as additional optimization parameters, widening the design space and enabling even stronger and more precisely engineered dispersion profiles.

Fabrication-induced variations in waveguide geometry can shift coupling ratios in directional couplers and alter the phase response of custom tapers, distorting the dispersion profile and resulting in higher insertion losses. To mitigate these effects, bent directional couplers \cite{el2024low} could serve as a low-loss, fabrication-tolerant alternative to regular directional couplers. Additionally, the optimization process itself also affects robustness as some minima of the error function \( L(\mathbf{W}) \) yield more robust final devices, making the resulting networks less sensitive to fabrication variations. This suggests that robustness can be improved also through complementary multi-objective approaches by guiding the optimizer towards specific dispersion targets in the intentional presence of fabrication-induced variations or even thermal fluctuations.

\section{Conclusion}

In conclusion, we proposed and experimentally demonstrated the design and optimization of arbitrary dispersion profiles using deep photonic networks with custom-designed MZIs. Our approach leverages the high degree of freedom in optimizable waveguide tapers in MZIs, enabling the realization of arbitrarily defined dispersion profiles with low insertion losses. Specifically, we demonstrated a proof-of-concept triangular dispersion profile over a 40 nm wavelength range (1.54 µm to 1.58 µm) with less than 0.7 dB insertion loss, significantly reducing losses compared to the literature. Additionally, we showed that our approach scales to larger networks with three or more ports, allowing independent control of more spectral features for spectrally more complicated or greater dispersion profiles. These capabilities in dispersion engineering highlight the potential of deep photonic networks for achieving application-specific phase/dispersion profiles, which can drive further innovations in pulse shaping and both classical and quantum information processing applications.

\section*{Acknowledgments}

This work was supported by H2020 Marie Skłodowska-Curie Actions (101032147); Türkiye Bilimsel ve Teknolojik Araştırma Kurumu (119E195)



\begin{thebibliography}{1}
\bibliographystyle{IEEEtran}
\bibitem{chenchip} G. F. R. Chen, K. Y. K. Ong, and D. T. H. Tan, "Chip-Scale Dispersion Compensation of High-Speed Data--Recent Progress and Future Perspectives," \emph{Laser \& Photonics Reviews}, p. 2400755, 2007.

\bibitem{dolgaleva2011integrated} K. Dolgaleva, A. Malacarne, P. Tannouri, L. A. Fernandes, J. R. Grenier, J. S. Aitchison, J. Azaña, R. Morandotti, P. R. Herman, and P. V. S. Marques, "Integrated optical temporal Fourier transformer based on a chirped Bragg grating waveguide," \emph{Opt. Lett.}, vol. 36, no. 22, pp. 4416--4418, 2011.

\bibitem{madsen1999integrated} C. K. Madsen, G. Lenz, A. J. Bruce, M. A. Cappuzzo, L. T. Gomez, and R. E. Scotti, "Integrated all-pass filters for tunable dispersion and dispersion slope compensation," \emph{IEEE Photon. Technol. Lett.}, vol. 11, no. 12, pp. 1623--1625, 1999.

\bibitem{vlasov2005active} Y. A. Vlasov, M. O'Boyle, H. F. Hamann, and S. J. McNab, "Active control of slow light on a chip with photonic crystal waveguides," \emph{Nature}, vol. 438, no. 7064, pp. 65--69, 2005.

\bibitem{wang2020integrated} X. Wang, L. Zhou, L. Lu, X. Wang, and J. Chen, "Integrated optical delay line based on a loopback arrayed waveguide grating for radio-frequency filtering," \emph{IEEE Photon. J.}, vol. 12, no. 3, pp. 1--11, 2020.

\bibitem{sahin2017large} E. Sahin, K. J. A. Ooi, C. E. Png, and D. T. H. Tan, "Large, scalable dispersion engineering using cladding-modulated Bragg gratings on a silicon chip," \emph{Appl. Phys. Lett.}, vol. 110, no. 16, 2017.

\bibitem{weiner2011ultrafast} A. M. Weiner, "Ultrafast optical pulse shaping: A tutorial review," \emph{Opt. Commun.}, vol. 284, no. 15, pp. 3669--3692, 2011.

\bibitem{mower2013high} J. Mower, Z. Zhang, P. Desjardins, C. Lee, J. H. Shapiro, and D. Englund, "High-dimensional quantum key distribution using dispersive optics," \emph{Phys. Rev. A}, vol. 87, no. 6, p. 062322, 2013.

\bibitem{madsen1998optical} C. K. Madsen and G. Lenz, "Optical all-pass filters for phase response design with applications for dispersion compensation," \emph{IEEE Photon. Technol. Lett.}, vol. 10, no. 7, pp. 994--996, 1998.

\bibitem{notaros2017programmable} J. Notaros, J. Mower, M. Heuck, C. Lupo, N. C. Harris, G. R. Steinbrecher, D. Bunandar, T. Baehr-Jones, M. Hochberg, S. Lloyd, \emph{et al.}, "Programmable dispersion on a photonic integrated circuit for classical and quantum applications," \emph{Opt. Express}, vol. 25, no. 18, pp. 21275--21285, 2017.

\bibitem{tahvili2013inp} S. Tahvili, S. Latkowski, B. Smalbrugge, X. J. M. Leijtens, P. J. Williams, M. J. Wale, J. Parra-Cetina, R. Maldonado-Basilio, P. Landais, M. K. Smit, \emph{et al.}, "InP-based integrated optical pulse shaper: demonstration of chirp compensation," \emph{IEEE Photon. Technol. Lett.}, vol. 25, no. 5, pp. 450--453, 2013.

\bibitem{moreira2016programmable} R. Moreira, S. Gundavarapu, and D. J. Blumenthal, "Programmable eye-opener lattice filter for multi-channel dispersion compensation using an integrated compact low-loss silicon nitride platform," \emph{Opt. Express}, vol. 24, no. 15, pp. 16732--16742, 2016.

\bibitem{bogaerts2020programmable} W. Bogaerts, D. Pérez, J. Capmany, D. A. B. Miller, J. Poon, D. Englund, F. Morichetti, and A. Melloni, "Programmable photonic circuits," \emph{Nature}, vol. 586, no. 7828, pp. 207--216, 2020.

\bibitem{ashtiani2022chip} F. Ashtiani, A. J. Geers, and F. Aflatouni, "An on-chip photonic deep neural network for image classification," \emph{Nature}, vol. 606, no. 7914, pp. 501--506, 2022.

\bibitem{najjar2024deep} A. N. Amiri, A. D. Vit, K. Gorgulu, and E. S. Magden, "Deep photonic network platform enabling arbitrary and broadband optical functionality," \emph{Nat. Commun.}, vol. 15, no. 1, p. 1432, 2024.

\bibitem{bradbury2018jax} J. Bradbury, R. Frostig, P. Hawkins, M. J. Johnson, C. Leary, D. Maclaurin, G. Necula, A. Paszke, J. VanderPlas, S. Wanderman-Milne, \emph{et al.}, "JAX: Composable transformations of Python+ NumPy programs, version 0.3.13," 2018.

\bibitem{trax} Google, "Trax: an end-to-end library for deep learning that focuses on clear code and speed," 2020. [Online]. Available: \url{https://github.com/google/trax}

\bibitem{vit2023silicon} A. D. Vit, K. Gorgulu, A. N. Amiri, and E. S. Magden, "Silicon photonics toolkit," \emph{Optica Open}, 2023.

\bibitem{el2024low} A. H. El-Saeed, A. Elshazly, H. Kobbi, R. Magdziak, G. Lepage, C. Marchese, J. R. Vaskasi, S. Bipul, D. Bode, M. E. Filipcic \emph{et al.}, "Low-loss silicon directional coupler with arbitrary coupling ratios for broadband wavelength operation based on bent waveguides," \emph{J. Lightw. Technol.}, 2024.


\end{thebibliography}
\end{document}